\documentclass[oldversion,referee]{aa}
\usepackage{times,graphicx,amssymb,lscape}
\usepackage{rotating}
\usepackage{natbib}
\bibpunct{(}{)}{;}{a}{}{,}  

\begin{document}

\title{Line intensity enhancements in stellar coronal X-ray spectra due to
opacity effects}

\author{S.J. Rose\inst{1}
        \and
        M. Matranga\inst{2,3}
	\and
        M. Mathioudakis\inst{2} 
        \and
        F.P. Keenan\inst{2}
	\and
	J.S. Wark\inst{4}
        }

\institute{Department of Physics, Imperial College, London SW7 2BZ, UK
\and
 Astrophysics Research Centre, School of Mathematics and Physics, Queen's University, Belfast, BT7~1NN,
Northern Ireland, U.K
\and
Harvard-Smithsonian Center for Astrophysics, MS-3, 60 Garden Street, Cambridge,
MA 02138, USA
\and
Department of Physics, Clarendon Laboratory, University of Oxford, 
Parks Road, Oxford, OX1 3PU, UK
}

\offprints{S.J. Rose, \email{s.rose@imperial.ac.uk}}

\date{Received date / Accepted date}

\abstract{
{\em Context.} The I(15.01 \AA)/I(16.78 \AA) emission line 
intensity ratio in Fe\,{\sc xvii} has been reported to deviate from its theoretical value in solar and stellar X-ray spectra. This is attributed to opacity in the 15.01 \AA\ line, leading to a reduction in its intensity, and was interpreted in terms of a geometry in which the emitters and absorbers are spatially distinct.\\
{\em Aims.} We study the I(15.01 \AA)/I(16.78 \AA) intensity ratio for the active cool dwarf
EV Lac, in both flare and quiescent spectra.  \\
{\em Methods.} The observations were obtained with the Reflection Grating Spectrometer on the XMM-Newton satellite. The emission measure distribution versus temperature reconstruction technique is used for our analysis. \\
{\em Results.} We find that the 15.01 \AA\ line exhibits a significant enhancement in intensity over the optically thin value. To our knowledge, this is the first time that such an enhancement has been detected on such a sound statistical basis. We interpret this enhancement in terms of a geometry in which the emitters and absorbers are not spatially distinct, and where the geometry is such that resonant pumping of the upper level has a greater effect on the observed line intensity than resonant absorption in the line-of-sight.
}

\authorrunning{S. J. Rose et al.}
\titlerunning{Line intensity enhancements in coronal X-ray spectra}

\maketitle 
 
\section{Introduction}
\label{intro}
Although opacity effects have been frequently detected in the chromospheric and transition region emission lines of lowly ionized species in solar and stellar sources (see, for example, Christian et al. 2006), this is not the case for coronal transitions in more highly ionized systems. In some respects, this is not surprising, as the optical depth at line centre scales directly with wavelength, and is inversely proportional to the square root of the plasma temperature (Mitchell \& Zemansky 1961). Coronal lines in highly charged ions will generally lie at short wavelengths (in the X-ray spectral region), and be formed at high values of plasma temperature. Schmelz et al. (1997) and Phillips et al. (1996) found some evidence of opacity in the Fe\,{\sc xvii}
15.01 \AA\ line from an analysis of solar active regions. However, no evidence of coronal opacity was found in the RS CVn binaries Capella (Phillips et al. 2001) and II Peg (Huenmoerder et al. 2001), nor in a sample of 26 late-type stars (Ness et al. 2003), although Testa et al. (2004) claim the detection of opacity in the O\,{\sc viii}
18.97 \AA\ emission line of II Peg. 

More recently, Matranga et al. (2005) have analysed a 50 ksec XMM-Newton observation of the late-type star AB Dor. They found that the observed emission 
line intensities were generally in very good agreement with theoretical values from an optically thin model. However, the Fe\,{\sc xvii} intensity ratio I(15.01 \AA)/I(16.78 \AA) was significantly smaller during the flare period of observation than the optically thin prediction. This was attributed to opacity in the 15.01 \AA\ transition, which has a much larger oscillator strength than the 16.78 \AA\ feature. There was no evidence of opacity in the quiescent X-ray spectrum of AB Dor. 

EV Lac (GJ 873) is a very active dM4.5e flare star at a distance of 5 pc. It exhibits spot-related quasi-periodic variations, and has been reported to have a solar-like activity cycle (Mavridis \& Avgoloupis 1986). EV Lac has one of the highest surface magnetic flux in the recent survey of low mass stars by Reiners \& Basri (2007). In the present paper we study the X-ray spectrum of EV Lac, with emphasis on the effects of resonant scattering, via the analysis
of Fe\,{\sc xvii} line intensity ratios. 

\section{Observations \& Analysis}
\label{Obser}
The observations of AB Dor reported by Matranga et al. (2005) were part of a larger dataset that also included the star EV Lac. Observations of these stars were made during both flare and quiescent periods and were obtained with the XMM-Newton Reflection Grating Spectrometers RGS1 and RGS2, which provide spectra with a resolution of 70--500 in the 0.2--2.5 keV range. Data were reduced with the XMM Science Analysis System (SAS) software v5.3.3. The RGS1 and RGS2 spectra are divided into 3400 channels with widths ranging from 0.007 to 0.014 \AA, and could not be summed directly as their wavelength grids do not coincide, and were therefore rebinned into a new grid with a width of 0.02 \AA\ 
for further analysis. A total line spread function was defined for the summed spectrum, which is the sum of the individual line spread functions weighted by the effective area (see Scelsi et al. 2004). The analysis of the RGS spectra is based on the plasma emission measure distribution versus temperature reconstruction technique, and was performed using the IDL-based software PINTofALE (Kashyap \& Drake 2000). 

In our analysis, the first step was the identification of the strongest lines in each spectrum, using the latest version (Version 5.2) of the CHIANTI atomic database (Dere et al. 1997; Landi et al. 2006).  The line spread function of the RGS instrument is characterized by extended wings, and multicomponent fitting involving the adjacent transitions was therefore necessary for most of the lines. In this case the total emissivity is the sum of the emissivities of the single lines, each multiplied by the corresponding element abundance. Lines rejected from our analysis include those whose inverse emissivity curves did not agree with those of other lines of the same ion. Weak lines, which are affected by large uncertainties and strong blending, were also excluded from the emission measure reconstruction. Further details of our analysis procedures may be found in Matranga et al. (2005).

\section{Results and Discussion}
\label{results}
The observed intensity ratios of the Fe\,{\sc xvii} 15.01\AA\ (3d $\rightarrow$ 2p) to 16.78\AA\ (3s $\rightarrow$ 2p) transitions are consistently smaller that those calculated by collisional-radiative models. In an effort to explain the discrepancy Laming et al. (2000) measured the ratio using the NIST EBIT and found it in agreement with the ratios predicted using a distorted wave collisional model. Similar experiments in the LLNL EBIT-II showed that resonant excitation can make a significant contribution to the 3s $\rightarrow$ 2p line intensities (Beiersdorfer et al. 2002). Additional processes therefore have to be taken into account. Doron \& Behar (2002) have shown that the inclusion of radiative recombination, dielectronic recombination, resonant excitation and inner-shell collisional ionization into a collisional-radiative model, can rectify some of these discrepancies. The Doron \& Behar (2002) calculations and are used in our analysis. We note that the most recent calculations of Landi \& Gu (2006) are in agreement with Doron \& Behar (2002).

 In Table 1 we present the observed I(15.01 \AA)/I(16.78 \AA) 
line intensity ratios for EV Lac during 4 different periods of observation, namely 1 quiescent and 3  flare, where the quoted errors in R$_{observed}$ correspond to 2$\sigma$. The predicted ratios for an optically thin plasma from the calculations of  Doron \& Behar (2002), R$_{theory}$, and the ratios generated from the differential emission measure (DEM) reconstruction, R$_{DEM}$, are also given for each case. Values quoted for R$_{theory}$ are for 2 different estimates of the plasma 
temperature (T$_{e}$),  with R$_{theory}$ = 1.75 calculated for T$_{e}$ = 7.2$\times$10$^{6}$ K, determined using the measured I(Fe\,{\sc xviii} 16.07 \AA)/I(Fe\,{\sc xvii} 16.78 \AA) intensity ratio in conjunction with the calculations of Phillips et al. (1997). By contrast, R$_{theory}$ = 1.85 is the theoretical ratio for T$_{e}$ = 
10$^{7}$ K, the temperature corresponding to the peak of the Fe\,{\sc xvii} 15.01 \AA\ line contribution function.

Table 1 shows several measured I(15.01 \AA)/I(16.78 \AA) ratios which are larger than the theoretical values, with the most pronounced example being for the quiescent phase of EV Lac, where the observed ratio is greater than the theoretical value by 2.6$\sigma$. (The estimated errors in the observations are smallest for the quiescent phase because the period of observation is largest in this case). This result implies enhancement in the 15.01 \AA\ line intensity, as opposed to the reduction in intensity which might normally
be expected for an optically thick transition. Our findings are confirmed in Fig. 1, which shows the EV Lac spectra for the four periods of observation. The observed spectra and those predicted assuming an optically thin plasma using the DEM reconstruction are markedly different only for the 15.01 \AA\ line during the quiescent period. For the quiescent spectrum, the 16.78 \AA\ line intensity is in good agreement with the optically thin model, whereas that of the 15.01 \AA\ line is more intense than the model prediction. We would like to emphasize that the DEM analysis of the flare and quiescent spectra included several strong lines in the 5-38\AA\ wavelength range. Below we give the lines used for the reconstruction of the quiescent spectrum together with their approximate temperatures of formation. Ne X (12.13\AA, logT $\sim$ 6.8), Ne IX (13.45\AA, logT $\sim$ 6.6), Fe XVIII (14.21\AA, \& 14.54\AA, logT $\sim$ 6.9), Fe XVII (17.10\AA, logT $\sim$ 6.7), O VIII (18.97\AA, logT $\sim$ 6.5), O VII (21.60\AA\, logT $\sim$ 6.3), N VII (24.78\AA, logT $\sim$ 6.3). The DEM distributions used in this work are presented in Fig. 2.

The statistical significance of the measurement of the enhancement depends on the 
degree to which we can safely assume that the EV Lac measurements are independent from 
one another and from the AB Dor observations (which showed no enhancement) obtained during the same series. 
We note that if they can be considered independent, the observed line ratio being larger than the theoretical value from the calculations of Doran and Behar (2002) by 
2.6$\sigma$ indicates a 99.5\%\ confidence level of our result being a true enhancement. If the theoretical value from the DEM reconstruction is used the value is reduced to 2.3$\sigma$ which indicates a 98.9\%\ confidence level. We proceed to outline a physical scenario for such an event. 

In Matranga et al. (2005), the interpretation of the line intensity ratio I(15.01 \AA)/I(16.78 \AA) being smaller than the optically thin value involved a model with a spatially separated emission and absorption region. To interpret this ratio being larger than the optically thin value, as found in the present work for the quiescent phase of EV Lac, we consider a single plasma that contains both emitter and absorber ions using the model developed by Kerr et al. (2004, 2005). Kerr et al. (2005) showed theoretically that, for spatially uniform conditions, the measured optically thick to optically thin line intensity ratio can be both smaller or (surprisingly) larger than the predicted optically thin value. Which of these occurs is dependent on two aspects of the plasma. The first is the overall geometry, which effectively determines the pumping of the upper state in the optically thick spectral line. The upper state of the transition is pumped by photons traversing the plasma at different angles and the characteristic length associated with this process is the mean chord length which is determined by the overall geometry of the plasma. Secondly, the orientation of the line-of-sight through the plasma determines the depletion by absorption of photons emitted by the upper state in the optically thick spectral line before they reach the observer. The characteristic length associated with this aspect is the average line of sight distance through the plasma. If the mean chord is longer than the average line of sight distance, then the pumping of the upper state is greater than the effect of depletion and the line intensity of the optically thick line is greater than the value assuming no optical depth. If the line of sight distance is greater than the mean chord length, the opposite is the case. In this way a line intensity ratio between an optically thick line and an optically thin line can, in principle, provide information on the plasma geometry and the orientation of the plasma to the observer.

A statistical study of coronal opacity in 26 late-type stars carried out 
by Ness et al. (2003) using Fe\,{\sc xvii}, revealed 
that EV Lac was the only object with significant line ratio anomalies. EV Lac 
has also been identified as one of the objects with significant optical depth in the sample of Testa et al. (2007). We emphasize that the previous studies were based on the intrpretation of time-averaged coronal spectra while in our analysis the flare and quiescent spectra have been separately analysed.

We do not attempt here to use the observed ratio in EV Lac to determine the geometry of the Fe\,{\sc xvii}
emitting region in the star. This is because we recognise that, in Kerr et al. (2005), many simplifying assumptions were made (such as uniform plasma conditions and simple planar, cylindrical or spherical geometries). Instead, we note that the observed line ratio being significantly 
larger than the theoretical value is consistent with the physics elucidated by Kerr et al. (2005). 

\section{Concluding Remarks}
\label{conclusions}

To extend our interpretation requires that we develop a detailed model of radiation transfer through realistic stellar coronal plasmas. In addition we plan to investigate the underlying physics in two further ways, both of which involve measuring line ratios from plasmas of known geometry. 
Firstly, we will use the Hinode and SOHO satellites to obtain time-series spectra
of optically thick and thin extreme-ultraviolet lines of Fe\,{\sc xv}
in solar active regions as they rotate 
from the solar disc centre to the limb. 
Simultaneous images in the Fe\,{\sc xv} 284\,\AA\ emission line 
will provide detailed
information on the active region morphology, and in particular will allow us to 
construct a 3-dimensional picture of its geometry as it rotates from the disc centre to the limb.
We will be able to search for and compare observed intensity enhancements in optically thick lines
with the predictions of our models for the known plasma geometry conditions. 
Secondly, we believe that we can produce a 
plasma of known geometry and measurable conditions within the laboratory, for which the 
geometrical dependence of an appropriate line ratio can be measured. A preliminary experiment 
has been designed by Rose et al. (2006), and further work on this and similar 
experiments is currently underway. The results of both solar and laboratory studies 
will be reported elsewhere. Finally, we note that further detailed studies of EV Lac 
would be of great benefit in determining the level of confidence that could be 
attributed to the observations of enhancement.  

\acknowledgements
FPK is grateful to AWE Aldermaston for the award of a William Penney Fellowship. CHIANTI is a collaborative project involving the Naval Research
Laboratory (USA), Rutherford Appleton Laboratory (UK), Mullard Space
Science Laboratory (UK), and the Universities of Florence (Italy) and
Cambridge (UK), and George Mason University (USA). The authors are grateful to Mr. David Mackenzie and Mr. Robert Green of Simon Langton Boys' School, Canterbury, for their help in the analysis.


\clearpage
\begin{table}
\begin{center}
\caption{Observed Fe\,{\sc xvii} line intensity ratios I(15.01 \AA)/I(16.78 \AA), R$_{observed}$, obtained using the XMM-Newton satellite for EV Lac during quiescent and flare periods. 
The theoretical line ratios, R$_{theory}$, from the calculations of Doron \& Behar (2002) and the ratios determined from the DEM spectrum, R$_{DEM}$ (red line in Fig 1), are also shown.}
\label{lines}
\begin{tabular}{lccc}
\hline
\hline
     &  R$_{observed}$     &  R$_{theory}$ & R$_{DEM}$\\
\hline
Quiescent &   2.50$\pm$0.50     &    1.75--1.85 & 1.93 \\          
Flare 1  &    2.38$\pm$1.25     &    1.75--1.85 & 1.77 \\
Flare 2  &    2.13$\pm$1.00     &    1.75--1.85 & 1.69 \\
Flare 3 & 1.64$\pm$0.64 & 1.75--1.85 & 1.89 \\
\end{tabular}
\end{center}
\end{table}

\clearpage

\begin{figure}
\begin{center}
\includegraphics[angle=0,width=12cm]{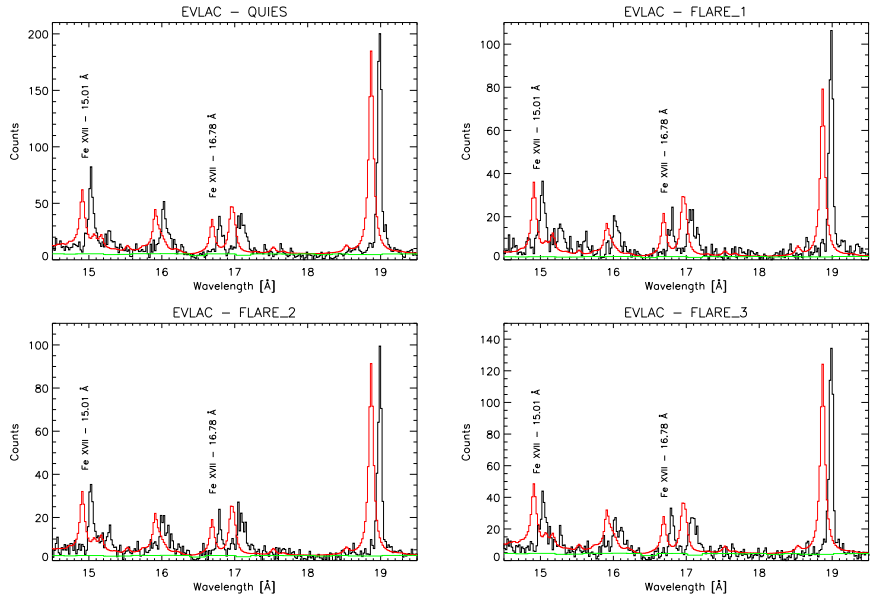}
\caption{\small Observed XMM-Newton spectrum (black line), the DEM reconstructed spectrum (itself based on atomic data from Doron \& Behar (2002)) (red line) and continuum (green line) for the star EV Lac during one quiescent and three flare periods. The wavelength grid of the predicted spectrum has been shifted by 0.1 \AA\ for clarity. For the quiescent period only, there is a clear 
increase of the Fe\,{\sc xvii}
15.01 \AA\ line intensity, which is not observed for the 16.78 \AA\ transition.}
\label{Spectrum}
\end{center}
\end{figure}

\clearpage

\begin{figure}
\begin{center}
\includegraphics[angle=0,width=12cm]{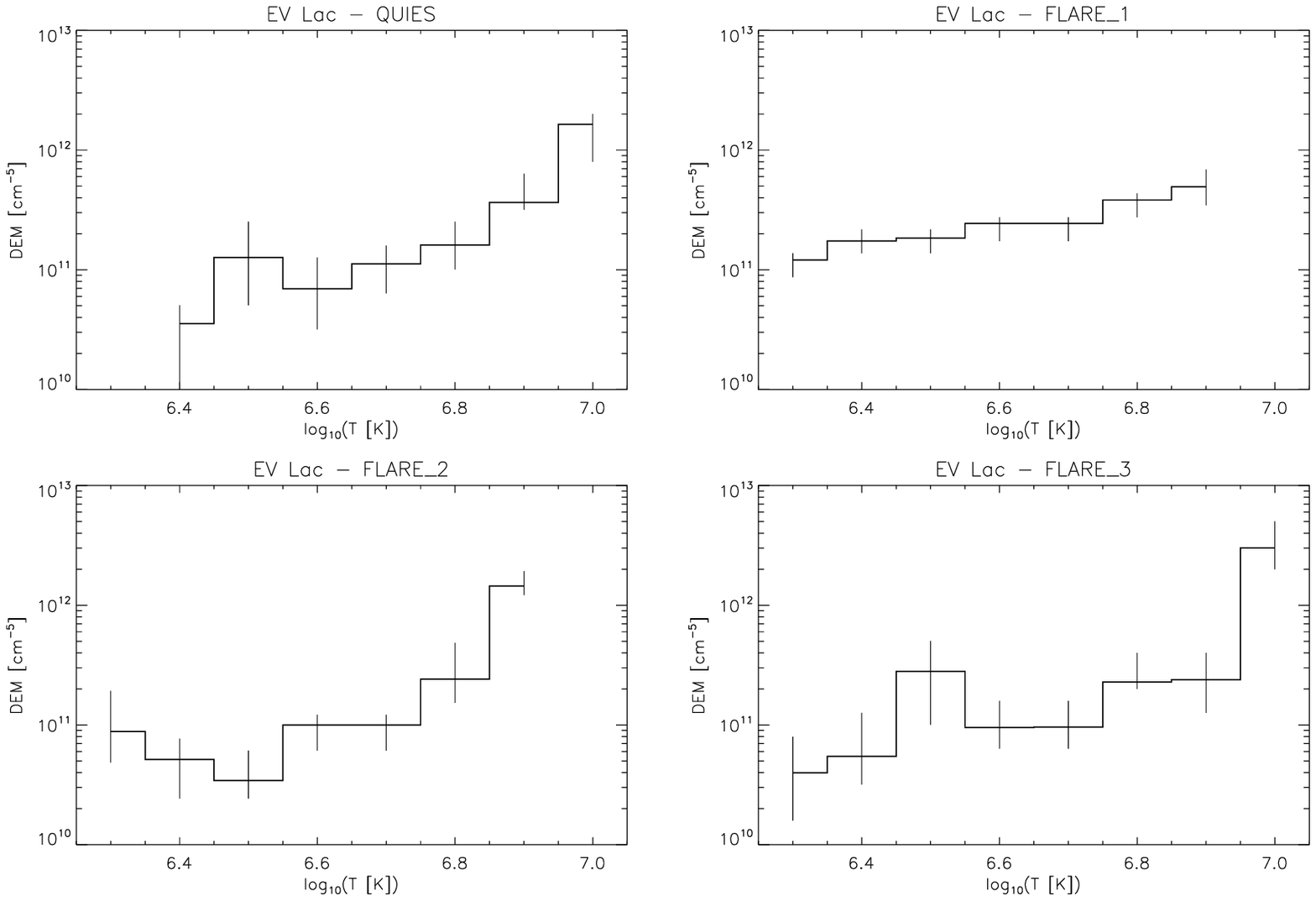}
\caption{\small The differential emission measure (DEM) distribution for the quiescent state and flares of EV Lac. This is defined as $DEM(T) = n_e^2 \frac{dV(T)}{dlogT}$ and is divided by 4 $\pi$ d$^2$ where $d$ is the stellar distance.}
\label{Spectrum}
\end{center}
\end{figure}

\clearpage

\bibliography{aa}
\bibliography{submit}
 
\bibitem[2002]{} 
Beiersdorfer, P., et al., 2002, ApJ, 576, L169 
\bibitem[2006]{}
Christian, D.J., Mathioudakis, M., Bloomfield, D.S., Dupuis, J., Keenan, F.P., Pollacco, D.L., Malina, R., 2006, A\&A, 454, 889
\bibitem[1997]{}
Dere, K.P., Landi, E., Mason, H.E., Monsignori Fossi, B.C., Young, P.R., 1997, A\&AS, 125, 149 
\bibitem[2002]{}
Doron, R. \& Behar, E., 2002, ApJ, 574, 518 
\bibitem[2001]{}
Huenemoerder, D.P., Canizares, C.R., Schulz, N.S., 2001, ApJ, 559, 1135  
\bibitem[2000]{}
Kashyap, V. \& Drake, J.J., 2000, ApJ, 503, 450
\bibitem[2004]{}
Kerr, F.M., Rose, S.J., Wark, J.S., Keenan, F.P., 2004, ApJ, 613, L181
\bibitem[2005]{}
Kerr, F.M., Rose, S.J., Wark, J.S., Keenan, F.P., 2005, ApJ, 629, 1091
\bibitem[2000]{}
Laming, J.M., et al. 2000, ApJ, 545, L161
\bibitem[2006]{}
Landi, E., Del Zanna, G., Young, P.R., Dere, K.P., Mason, H.E., Landini, M., 2006, ApJS, 162, 261 
\bibitem[2006]{}
Landi, E., Gu, M.F., 2006, ApJ, 640, 1171
\bibitem[2005]{}
Matranga, M., Mathioudakis, M., Kay, H.R.M., Keenan, F.P., 2005, ApJ, 621, L125 
\bibitem[1986]{}
Mavridis, L.N., Avgoloupis, S., 1986, A\&A 154, 171
\bibitem[1961]{}
Mitchell, A.C.G. \& Zemansky, M.W., 1961, Resonance Radiation and Excited Atoms (Cambridge: Cambridge University Press).  
\bibitem[2003]{}
Ness, J.-U., Schmitt, J.H.M.M., Audard, M., Güdel, M., Mewe, R., 2003, A\&A, 407, 347 
\bibitem[1996]{}
Phillips, K.J.H., Greer, C.J., Bhatia, A.K, R., Keenan, F.P., 1996, ApJ, 469, L57 
\bibitem[1997]{}
Phillips, K.J.H., Greer, C.J., Bhatia, A.K., Coffey, I.H., Barnsley, R., Keenan, F.P., 1997, A\&A, 324, 381 
\bibitem[2001]{}
Phillips, K.J.H., Mathioudakis, M., Huenemoerder, D.P., Williams, D.R., Phillips, M.E., Keenan, F.P., 2001, MNRAS, 325, 1500 
\bibitem[2007]{}
Reiners, A., Basri, G., 2007, ApJ, 656, 1121
\bibitem[2006]{}
Rose, S.J., Kerr, F.M., Wark, J.S., Keenan, F.P., 2006, Plasma Physics Department Annual Report AWE, Aldermaston, 119 
\bibitem[1997]{}
Schmelz, J.T., Saba, J.L.R., Chauvin, J.C., Strong, K.T., 1997, ApJ, 477, 509 
\bibitem[2004]{}
Scelsi, L., Maggio, A., Peres, G., Gondoin, P., 2004, A\&A, 413, 643 
\bibitem[2004]{}
Testa, P., Drake, J.J., Peres, G., DeLuca, E.E., 2004, ApJ, 609, L79 
\bibitem[2007]{}
Testa, P., Drake, J.J., Peres, G., Huenemoerder, D.P., 2007, ApJ, 665, 1349


\end{document}